\documentstyle[preprint,aps]{revtex}

\def\beq{\begin{equation}}
\def\eeq{\end{equation}}
\def\be{\begin{eqnarray}}
\def\ee{\end{eqnarray}}
\def\ci{\cite}
\def\bi{\bibitem}

\def\PkE{P(|{\bf k}|,E)}
\def\magk{|{\bf k}|}

\begin{document}

 
\draft

\title{On the behaviour of the nuclear spectral function at high momentum 
and removal energy} 

\author
{ O. Benhar$^{1}$\cite{byline},  S. Fantoni${^2}$, G.I. Lykasov${^3}$ }

\address
{ $^1$ Jefferson Laboratory, Newport News, VA 23606, USA \\
and \\ Department of Physics, Old Dominion University,
Norfolk, VA 23529, USA \\
$^2$ Interdisciplinary Laboratory for Advanced Studies (ILAS)\\
and \\INFN, Sezione di Trieste. I-30014 Trieste, Italy\\
$^3$ Joint Institute for Nuclear Research, Dubna 141980, Moscow
Region, Russia }

\date{\today}

\maketitle

\begin{abstract}

We propose a procedure to extrapolate the nuclear spectral 
function $\PkE$ obtained from nonrelativistic many-body 
theory to large values of three-momentum and removal energy. 
Our approach is based on phenomenological information 
extracted from both soft hadron-nucleus interactions, 
in the regime where the proton inclusive spectrum is dictated by 
Regge asymptotic, and deep-inelastic lepton-nucleus collisions. 
The extrapolated $\PkE$ is used to compute the semi-inclusive 
spectra of backward protons produced in electron-nucleus
scattering.  

\end{abstract}
 

The knowledge of the spectral function $\PkE$, giving 
the probability to find a nucleon of momentum $\magk$ and 
removal energy $E$ inside a nucleus, is a prerequisite for 
the theoretical description of a number of reactions
involving nuclear targets. 
For both infinite nuclear matter and light nuclei, with mass number 
A$\leq$ 4, it is possible to carry out accurate calculations of 
$\PkE$ starting from a realistic nuclear hamiltonian, fitted to 
nucleon-nucleon scattering data and to the properties of 
few-nucleon bound states \ci{bf1,bp}. In the case of medium-heavy nuclei, 
quantitative estimates of the spectral function and its energy
integral, the momentum distribution $n(\magk)$, can also be obtained, 
 using the local density approximation \ci{bf2}. 

The spectral functions resulting from realistic many-body calculations
contain information on both the nuclear mean field (at low $\magk$ and $E$), 
and short-range nucleon-nucleon correlations (at high $\magk$ and $E$).
Tipically, only about 70\% of the nucleons are in the
states of low  $\magk$ and low $E$, that can be described by mean field
single-particle wave functions, while the remaining 30\% are in a 
correlated state with another nucleon, mainly on account of the 
one-pion-exchange tensor force and the short-range
repulsion of the nucleon-nucleon interaction.
The mean field ($P_0(\magk,E)$) and correlation ($P_B(\magk,E)$) contributions
to the spectral function can be singled out rewriting the full $\PkE$ 
in the form
\beq
\PkE = P_0(\magk,E) + P_B(\magk,E)\ .
\eeq

Strongly correlated nucleons play a very important role in many 
processes.
For example, the production of fast backward hadrons in semi-inclusive
lepton-nucleus reactions, in the kinematical region forbidden to
scattering off a free nucleon, is mostly due to nucleon-nucleon 
correlations \ci{bfl1,bfl3}. 
While many-body calculations tipically provide a description 
of the correlation tail of $\PkE$ up to $k_{max} \sim$ .7 GeV/c and 
$E_{max} \sim$ .6 GeV, the theoretical analysis of the
available spectra of leptoproduced backward hadrons carrying large
momentum \ci{slac} requires the knowledge of the nuclear spectral function 
at larger values of $\magk$ and $E$ \ci{bfl1}.
In this note, we propose a simple phenomenological procedure that 
allows to extrapolate the correlation contribution, $P_B(\magk,E)$,  
beyond the region covered by nuclear many-body theory. 

Let us consider a process in which a four-momentum $q \equiv (\nu,{\bf q})$ 
is transferred 
to a nuclear target. Our starting point is the relationship between the 
nuclear spectral function and the function $f_A(z)$, yielding the
distribution of the nucleons in the target as a function of
the relativistic invariant variable $z$, defined as 
\beq
z = \frac{M_A}{m}\frac{(kq)}{(P_Aq)}\ .
\label{def:z}
\eeq
In the above equation $m$ is the nucleon mass, whereas $P_A \equiv (M_A,0)$, 
$M_A$ being the target 
mass, and $k \equiv (k_0,{\bf k})$ denote the initial nuclear and nucleon 
four-momentum in the target rest frame, respectively. 
The distribution function $f_A(z)$ can be written in a general fashion as
\beq
f_A(z) = z \int d^4k\ S(k)\ 
\delta \left( z - \frac{M_A}{m}\frac{(kq)}{(P_Aq)} \right)\ ,
\label{1}
\eeq
where $S(k)$ is the relativistic function describing the nuclear vertex
with an outgoing nucleon of four-momentum $k$ (see, e.g., ref.\ci{fs}). 
$S(k)$ can be approximated by 
the nonrelativistic spectral function according to \ci{bfl2}
\beq
S(k)=\left(\frac{m}{k_0}\right) P(\magk,E)\ , 
\label{2}
\eeq
with
\beq
k_0 = M_A - \left[\left(M_A-m+E\right)^2+{\magk}^2\right]^{1/2}. 
\label{3}
\eeq
Note that the above definition implies that $f_A(z)$ also depends upon 
$Q^2 = -q^2$, as pointed out in \ci{bfl2}. However, in 
the following we will 
be assuming that the Bjorken limit ($Q^2, \nu \rightarrow \infty$) 
is applicable, so that $z$ can be related to the light-cone
component of the nucleon four-momentum through 
$z = (k^+/m) = (k_0 - k_z)/m$ ($k_z = ({\bf k}{\bf q})/\magk |{\bf q}|$) and 
the $Q^2$-dependence of $f_A(z)$ disappears.

Substituting eq.(\ref{2}) into eq.(\ref{1}) we can rewrite 
$f_A(z)$ in terms of $\PkE$ as
\beq
f_A(z) = 2 \pi m z \int_{E_{min}}^{\sqrt{s}-M_A} 
dE\int_{k_{min(z,E)}}^{\infty}
d\magk \magk\ \left( \frac{m}{k_0}\right) \PkE
\label{4}
\eeq
where $E_{min}$ is the minimum energy required to remove a nucleon from the 
target nucleus, $s = (P_A + q)^2$ and 
 (for simplicity, we work in the infinite nuclear matter limit, 
 $M_A \rightarrow \infty$, in which the kinetic energy of the 
recoiling nucleus becomes vanishingly small)
\beq
k_{min}(z,E)=|m(1-z)-E|\ .
\eeq
The above equations imply that the large $z$ behaviuor
of $f_A(z)$ is dictated by the high $\magk$ tail of $\PkE$, 
i.e. from its correlation part $P_B(\magk,E)$. For
example, at $z > 1.7$ only momenta larger than $ k_{max} \sim$ .7 GeV/c, 
contribute to the integral of eq.(\ref{4}). 

In ref.\ci{bfl1} we have developed a procedure 
to evaluate the asymptotic behavior of $f_A(z)$ without making use of the
nuclear spectral function. Within this approach, based on ideas originally 
proposed in refs.\ci{ekl1,ekl2}, $f_A(z>1)$ can be written as a sum of  
distributions $f_n(z/n)$, describing clusters of $n\  (n \geq 2)$ strongly 
correlated nucleons, whose asymptotic behavior as $z \rightarrow n$ can be 
evaluated for any value of $Q^2$. 
We will show that using the large $z$ behavior 
of $f_A(z)$ resulting from the approach of ref. \ci{bfl1} and 
eq.(\ref{4}) one can extract information on 
the behavior of  $\PkE$ at very large $\magk$ and E. 

For any given $\magk$, the nuclear 
matter $P_B(\magk,E)$ of ref. \ci{bf1} exhibits a bump  
located at $E \sim E_k = \sqrt{\magk^2 + m^2} -m$, the energy needed to 
remove a nucleon of momentum ${\bf k}$ belonging to a strongly correlated 
pair of vanishing total momentum, whose width increases with $\magk$. 
A remarkably accurate fit to the behaviour of the correlation contribution 
to the calculated spectral function has been obtained in ref.\cite{cls90}, 
where $P_B(\magk,E)$ has been written in the form
\beq
P_B(\magk,E) = n_B(\magk)F(\magk,E)\ ,
\label{appr:pke}
\eeq
with
\beq
n_B(\magk) = \int_{E_{thr}}^\infty dE\ P_B(\magk,E)\ ,
\label{nB}
\eeq
$E_{thr}$ being the minimum energy required to remove a nucleon pair.
 The function $F(\magk,E)$ is defined as
\beq
F(\magk,E) = N_k\  {\rm exp} \left\{ 
\frac{ \left[\ \sqrt{m(E-E_{thr})} - \sqrt{m(E_k-E_{thr})}\ \right]^2}
{2 \ \sigma_k^2} \right\} \ ,
\label{FE}
\eeq
where $\sigma_k$ is related to the width $\Gamma_k$ through
\beq
\Gamma_k = 4\ \sigma_k\ \sqrt{(2\ ln2)(E_k/m)} = \langle E_B \rangle + E_k\ ,
\label{def:gamma}
\eeq
$\langle E_B \rangle$ being the average removal energy associated with
$P_B(\magk,E)$ (using the spectral function of ref.\cite{bf1} one 
finds $\langle E_B \rangle \sim$ 40 MeV).
The normalization constant $N_k$ is chosen in such a way as to fulfill
 the sum rule
\beq
\int_{E_{thr}}^\infty dE\ F(\magk,E) = 1\ ,
\eeq
which in turn guarantees the overall normalization of $P_B(\magk,E)$. 

Our extrapolation procedure is based on the assumption that at very large
values of $\magk$ and $E$ the energy dependence of $P_B(\magk,E)$ is still
dominated by the contribution associated with the removal of a correlated 
nucleon pair, and can be described by $F(\magk,E)$ of eq.(\ref{FE}). As a
consequence, the extrapolation of $\PkE$ reduces to the extrapolation 
of $n_B(\magk)$.

Substitution of eq.(\ref{appr:pke}) into eq.(\ref{4}) leads (after
inversion of the integration order) to the
following expression for the correlation contribution to $f_A(z)$ at
large $z$: \beq
f_A^B(z) = 2 \pi m z \int_{k_{min}(z,E_{thr})}^\infty d\magk\ \Phi(\magk)\ ,
\label{invert}
\eeq
with
\beq
\Phi(\magk) = \magk\  n_B(\magk)\ \int_{E_{thr}}^{\magk - m(z-1)} dE 
\left( \frac{m}{k_0}\right) F(\magk,E)\ ,
\eeq
implying in turn
\beq
\phi^\prime(z) = \frac{d}{dz} \left( \frac{f_A^B(z)}{z} \right) =
- 2 \pi m^2  \Psi(z)\ ,
\label{def:phi}
\eeq
where
\beq
\Psi(z)= m \int_{E_{thr}}^\infty dE\ \frac{E+m(z-1)}{m-E}\  
n_B(E+m(z-1))\ F(E+m(z-1),E)\ .
\label{def:Psi}
\eeq

In general, it is not possible to invert eq.(\ref{def:phi}) to 
obtain $n_B(\magk)$ in terms of $\phi^\prime(z)$, i.e. of $f_A^B(z)$ 
and its first derivative. 
We have circumvented this problem using a parametrized $n_B(\magk)$ in 
eq.(\ref{def:Psi}) and fixing the values of the parameters in 
such a way as to reproduce the left hand side of eq.(\ref{def:phi}), 
evaluated according to ref.\cite{bfl1}. 

At $k_F < \magk < k_{max}$, $k_F \sim .25$ GeV/c being the Fermi momentum, 
the behavior of the momentum distribution of ref.\cite{bf1} 
is nearly exponential, and can be accurately approximated using
\beq
n_0(\magk) = G_1 {\rm exp}[-(B_1\magk)^\alpha]\ ,
\label{app:nk}
\eeq
with $G_1$ = 3.30 (GeV/c)$^{-3}$, $B_1$ = 6.2 (GeV/c)$^{-1}$ and 
$\alpha$ = 1.14. 

The most natural choice is to extend the parametrization of eq.(\ref{app:nk})
to larger $\magk$ and use it to calculate $\Psi(z)$ from eq.(\ref{def:Psi}). 
It should be noted, however, that the integrand in eq.(\ref{def:Psi}) has a 
singularity at $E=m$. This problem has been taken care of inserting a cutoff
to guarantee that $n_B(\magk)$ vanish at $\magk = mz$. 
The parametrization of $n_B(\magk)$ providing the best fit to $f_A^B(z)$ is
\beq
n_B(\magk) = {\rm exp}\left[- \beta\left( \frac{\magk - k_0}{\Lambda - \magk}
\right)\right]\ 
n_0(\magk)\ ,
\label{def:parnk}
\eeq
where $\beta = .001$, $k_0 \sim k_{max}$, 
$\Lambda = mz_0 = m + k_0 + E_{thr}$
and 
\beq
n_0(\magk) = G_1 {\rm exp}[-(B_1\magk)^\alpha] + 
G_2 {\rm exp}(-B_2\magk)\ ,
\label{def:parnk0}
\eeq
with $G_1$ = 2.90 (GeV/c)$^{-3}$, $B_1$ = 6.08 (GeV/c)$^{-1}$, 
$G_2$ = -21.2 (GeV/c)$^{-3}$ and $B_2$ = 26.7 (GeV/c)$^{-1}$. 
With the 
above choice of $n_B(\magk)$, $\Psi(z)$ is well defined and can be calculated
using eq.(\ref{def:Psi}) for any $z > z_0$. 

The results of numerical calculation show that the use of eq.(\ref{def:parnk})
with $n_0(\magk)$ given by eq.(\ref{def:parnk0}) leads to a remarkably good fit to 
$f_A^B(z)$ at $z > 1.7$, the corresponding $\chi^2$ being $\sim$ 0.009.

A more straightforward procedure to extract $n_B(\magk)$ from 
eq.(\ref{def:phi}) can be obtained making the rather drastic assumption that in 
eq.(\ref{def:Psi}) the function $F(\magk,E)$ can be
replaced by a $\delta$-function:
\beq
F(E+m(z-1),E) = \delta(E-(\sqrt{(E+m(z-1))^2+m^2}-m))\ .
\label{delta}
\eeq
Using eq.(\ref{delta}) the $E$ integration in eq.(\ref{def:Psi}) can 
be readily carried out and substitution of the result into eq.(\ref{def:phi})
leads to: 
\beq
n_B(k_s) = \frac{1}{2\pi m^2k_s}\ \frac{(3-z^2)(2-z)}{(2-z)^2+1}\ 
\phi^\prime(z)\ ,
\label{nk:delta}
\eeq
with $k_s=m(z-1)(z-3)/(2(z-2))$. Eq.(\ref{nk:delta}) 
gives $n_B(\magk)$
in terms of $\phi^\prime(z)$ for any values of $z$ in the range 
$1 < z < \sqrt{3}$.

Let us now focus on the calculation of the left hand side of 
eq.(\ref{def:phi}). According to ref. \ci{bfl1}, at large values of its 
argument $f_A^B(z)$ can be written as a sum, whose terms describe the 
contributions associated with strongly correlated $n$-nucleon clusters: 
\beq
f_A^B(z)=\sum_{n=2}^A f_n\left(\frac{z}{n}\right)\ , 
\label{12}
\eeq
the $n$-th term in the sum being defined for $1 < z < n$. 
Within this approach
the calculation of $f_A^B(z)$ reduces to the calculation of the relevant
$f_n(z/n)$'s, corresponding to the lowest values of $n$ (tipically $n$ =
2 and 3). 
The analysis of nuclear fragmentation in hadron-nucleus collisions
carried out in refs. \ci{ekl1,ekl2} shows that, at low $Q^2$, the 
distribution of 
colorless three-quark systems in a 3$n$-quark cluster, $T_n(z/n)$, 
exhibits true Regge asymptotic behavior as $z \rightarrow n$. 
The results of refs. \ci{ekl1,ekl2} provide a satisfactory description
of the inclusive spectra of high-momentum protons and mesons 
emitted backward in proton-nucleus collisions. However, the sizeable
$Q^2$-dependence exhibited by $f_A(z)$ at low $Q^2$ \ci{bfl2} suggests
that the nonpertubative $Q^2$-dependence of $T_n(z/n)$ has to 
be carefully taken into account. Starting from the small $Q^2$ 
behavior, which can be described within the
framework of Regge theory, the asymptotic 
$T_n(z/n)$ at $z \rightarrow n$ and large $Q^2$
 can be obtained 
from the distribution of valence quarks inside  
a cluster of $n$ strongly correlated nucleons, which can 
in turn be written in terms of
 the relativistic invariant phase-space volume 
available to a quark in a nucleon \ci{bfl1}. 

The function $T_n(z/n)$ can be interpreted as the
distributions of {\em effective nucleons}
within a strongly correlated cluster. Therefore, assuming that the valence
quark distribution inside these {\em effective nucleons} is the
same as in ordinary nucleons, the quantity 
${\widetilde T}_n(z/n) = w_nT_n(z/n)$,
 $w_n$ being the probability of finding an $n$-body cluster \cite{bfl1}, 
can be identified with $f_n(z/n)$ 
of eq.(\ref{12}). The validity of this approximation, which allows to 
effectively take into account nuclear excitations, is supported by the results 
of calculations of the valence quark distributions
inside nucleons and baryonic resonances \ci{buch,neud}.

Writing the distribution of valence quarks inside a nucleon at large $Q^2$
in the form
\beq
f_{q_v}^N(z) = C_N z^{a_N} (1-z)^{b_N}\ ,
\label{13}
\eeq
where $C_N$ is a normalization constant, 
$a_N = -\alpha_R(0) = 1/2$, $\alpha_R(0)$ being the intercept of the
Regge trajectory, and  $b_N \sim 2.8 - 3.2$, the behavior of
${\widetilde T}_n(z/n)$ as $z \rightarrow n$ can be obtained in closed form 
\ci{kaid1}. Under the assumption ${\widetilde T}_n(z/n) = f_n(z/n)$ 
we can use this result and write: 
\beq
f_n\left(\frac{z}{n}\right) = D_n \left(\frac{z}{n}\right)^{A_n}
\left[ 1 - \left(\frac{z}{n}\right) \right]^{g_n}\  ,
\label{14}
\eeq
with $g_n=(a_N + b_N + 2)(n-1) - 1$ and $A_n=a_N + b_N +1$, whereas the
coefficients $D_n$ can be obtained from the quark distribution
in the $n$-nucleon cluster, evaluated as in ref. \ci{bfl1}. The details 
of the derivation of eq.(\ref{14}) are given in the Appendix.

Substituting the $f_n(z/n)$'s  with $n$ = 2,3 and 4, calculated from 
eq.(\ref{14}), into eq.(\ref{12}), one can easily obtain both $f_A^B(z)$ 
and its first derivative at $z < 4$. The resulting $\phi^\prime$, defined 
as in eq.(\ref{def:phi}), can be written:
\beq
\phi^\prime (z) = I_2 + I_3 + I_4\ ,
\label{g1}
\eeq
where $I_n(z)$ is defined for $z < n$ and
\beq
I_2 = \frac{1}{2} D_2 \left( \frac{z}{2} \right)^{A_2-2}
\left[ 1- \left( \frac{z}{2} \right) \right]^{g_2-1}
\left\{ \left( A_2 - 1 \right) 
\left[ 1 - \left( \frac{z}{2} \right) \right] - 
g_2 \left( \frac{z}{2} \right) \right\}\ , 
\label{g2}
\eeq
\beq          
I_3 = \frac{1}{3} D_3 \left( \frac{z}{3} \right)^{A_3-2}
\left[ 1- \left( \frac{z}{3} \right) \right]^{A_3-2}
\left[ 1- \left( \frac{z}{3} \right) \right]^{g_3-1}
\left\{ \left( A_3 - 1 \right)
\left[ 1 - \left( \frac{z}{3} \right) \right] -
g_3 \left( \frac{z}{3} \right) \right\}
\label{g3}
\eeq
\beq  
I_4 = \frac{1}{4}D_4 \left(\frac{z}{4}\right)^{A_4-2}
\left[ 1- \left(\frac{z}{4}\right) \right]^{g_4-1}
\left\{\left( A_4 - 1 \right)\left[ 1 - \left(\frac{z}{4}\right) \right] 
- g_4\left(\frac{z}{4}\right) \right\}\ .
\label{15}
\eeq 
Finally, the momentum distribution
at large $\magk$ can be obtained from $\phi^\prime(z)$ defined by 
eqs.(\ref{g1})-(\ref{15}) using either eq.(\ref{def:phi}) and the 
fitting procedure or the $\delta$-function ansatz leading to 
eq.(\ref{nk:delta}). 

Fig. \ref{fig1} shows the nuclear matter momentum distribution 
$n(\magk)$ of 
ref. \ci{bf1}, calculated for $\magk \le$ .8 GeV/c (solid line), 
together with the values obtained from the fit to $\phi^\prime(z)$
at $\magk >$ .8 GeV/c (diamonds). It appears that the extrapolated tail of 
$n(\magk)$ is close to a simple exponential dependence. The results
obtained using the $\delta$-function approximation, which provides 
an upper bound to $n_B(\magk)$ at $\magk >$ .8 GeV/ci are also shown in
fig. 1 (dashed line). It is apparent that the energy spread of the 
strength is very important and cannot be disregarded. 

To test the extrapolated momentum distribution, we have calculated 
the kinetic energy spectrum of protons emitted at backward angle
in the semi-inclusive $e + A \rightarrow e^\prime + p + X$ reaction, 
in the kinematics of the SLAC data of ref. \ci{slac}, using \ci{bfl1}
\beq
\rho_{e A \rightarrow e^\prime p X}(x,Q^2,z) = 
\frac{I_{eN}}{I_{eA}} n_B\left[\magk(z)\right]F_2^N(x,Q^2)\ .
\eeq
In the above equation, 
$F_2^N(x,Q^2)$, $x$ being the Bjorken scaling variable, is the
nucleon structure function, while $I_{eN}$ and $I_{eA}$ denote the fluxes 
associated with scattering off an isolated nucleon and the nuclear 
target, respectively. In fig. \ref{fig2} the spectrum obtained using the
nucleon momentum distribution of ref. \ci{bf1} at $k < k_{max} \sim .8$ GeV/c 
(see ref.\ci{bfl3}) and setting $n_B(\magk > k_{max})= 0$
is compared to that obtained using $n(\magk > k_{max})$ obtained from
the fit to $f_A^B(z)$ at $z \ge 1.7$. It appears that 
the extrapolated $n_B(\magk)$, needed to describe the spectrum at large
proton energy ($T \ge .3$ GeV), also provides a much better fit to the 
data at lower T.

In conclusion, we have proposed a simple phenomenological procedure that
allows to extrapolate the nuclear matter spectral function beyond the
region described by nonrelativistic many-body calculations. The main
assumption needed to extract $n_B(\magk)$ from the relationship between 
the spectral function
and the distribution function $f_A(z)$, i.e. the assumption that 
$P_B(\magk,E)$, exhibits the $E$-dependence associated with the removal of
a nucleon belonging to a strongly correlated pair, appears to be adequate
in the range of momentum and removal energy relevant to our analysis. 
The importance of the description of the $E$-dependence is 
clearly shown by the fact that the oversimplified $\delta$-function ansatz 
leads to a severely overestimated momentum distribution. 
The numerical results shown in figs. \ref{fig1} and \ref{fig2} suggest 
that our approach can be used to  
quantitatively investigate reactions sensitive to the very high momentum
components of the nuclear wave function. 

\acknowledgements

This work has been encouraged and supported by the Russian 
Foundation of Fundamental Research. We gratefully acknowledge
many helpful discussions with A. Fabrocini. 

\appendix
\section{} 
 
In ref. \ci{bfl1} it has been shown that, assuming that the 
distribution of valence quarks inside an isolated nucleon can be 
described by
\beq
f_{q_v}^N(z)  \sim z^{a_N} (1-z)^{b_N}\ ,
\label{singleN}
\eeq
the corresponding distribution inside a 
strongly correlated $n$-nucleon cluster, obtained from the overlap of
the phase-space volumes available to a quark inside a nucleon, takes 
the form 
\beq
f_{q_v}^{(n)}\left(\frac{z}{n}\right) = 
{C}_n \left(\frac{z}{n}\right)^{a_N} 
\left[1 - \left(\frac{z}{n}\right)\right]^{[b_N+(a_N+b_N+2)(n-1)]}\ ,
\label{A1}
\eeq
where the coefficient ${C}_n$ can be written in terms of 
beta-functions. 

The function $f_{q_v}^{(n)}(z/n)$ 
can also be given in terms of 
the distribution of colorless three-quark systems within a 
3$n$-quark cluster, ${\widetilde T}_n(z^\prime)$, using Mellin's convolution 
formula:
\be
f_{q_v}^{(n)}\left(\frac{z}{n}\right) = 
\int_{(z/n)}^1 {\widetilde T}_n(z^\prime) f_{q_v}^{3Q}
\left(\frac{1}{n}\frac{z}{z^\prime}\right)
\frac{dz^\prime}{z^\prime}\ ,
\label{A2}
\ee
where $f_{q_v}^{3Q}(z/nz^\prime)$ 
denotes the distribution 
of valence quarks inside the colorless three-quark system. 

Assuming that at large $z/(nz^\prime)$ $f_{q_v}^{3Q}(z/nz^\prime)$ 
 is approximately the same as the distribution inside a
nucleon, e.g. assuming $f_{q_v}^{3Q} \sim f_{q_v}^N$, and using 
eq.(\ref{singleN}), eq.(\ref{A2}) can be inverted to obtain  
${\widetilde T}_n(z^\prime)$ in the form:
\be
{\widetilde T}_n(z) = D_n z^{A_n} (1 - z)^{g_n}\ ,
\label{A3}
\ee
with $D_n$, $A_n$ and $g_n$ given in terms of the constants
$C_n$, $a_N$ and $b_N$ appearing in eq.({\ref{A1}).

Within this approximation eq.(\ref{A3}) reduces to eq.(\ref{14}), since
${\widetilde T}_n$ can be identifyed with $f_n$, which can in turn 
be interpreted as the distribution of colorless
three-quark systems inside of a strongly correlated $n$-nucleon 
cluster.

\begin{figure}
\caption{The nucleon momentum distribution in infinite nuclear matter.
The solid line corresponds to the $n_B(\magk)$ of ref. \protect\cite{bf1}, 
whereas the diamonds show the large $\magk$ extrapolation 
parametrized according to 
eqs.(\protect\ref{def:parnk})-(\protect\ref{def:parnk0}).
The dashed line has been obtained using the $\delta$-function
approximation of eq.(\protect\ref{delta}).}
 
\label{fig1}
\end{figure}

\begin{figure} 
\caption{Kinetic energy spectra of protons emitted at 
angle $\Theta$ in the
semi-inclusive reaction $e + CO \rightarrow e^\prime + p + X$. The dashed 
curve has been obtained using the nucleon momentum distribution 
of ref.\protect\cite{bf1} at $\magk < k_{max} \sim $ 0.8 GeV/c and
assuming $n_B(\magk > k_{max}) = 0$, see ref.\protect\cite{bfl3}, 
whereas the solid line shows
the results obtained using the large $\magk$ extrapolation, 
parametrized according to
eqs.(\protect\ref{def:parnk})-(\protect\ref{def:parnk0}), 
at $\magk > k_{max}$.
The experimental data are taken from ref.\protect\ci{slac}.}
\label{fig2}
\end{figure}

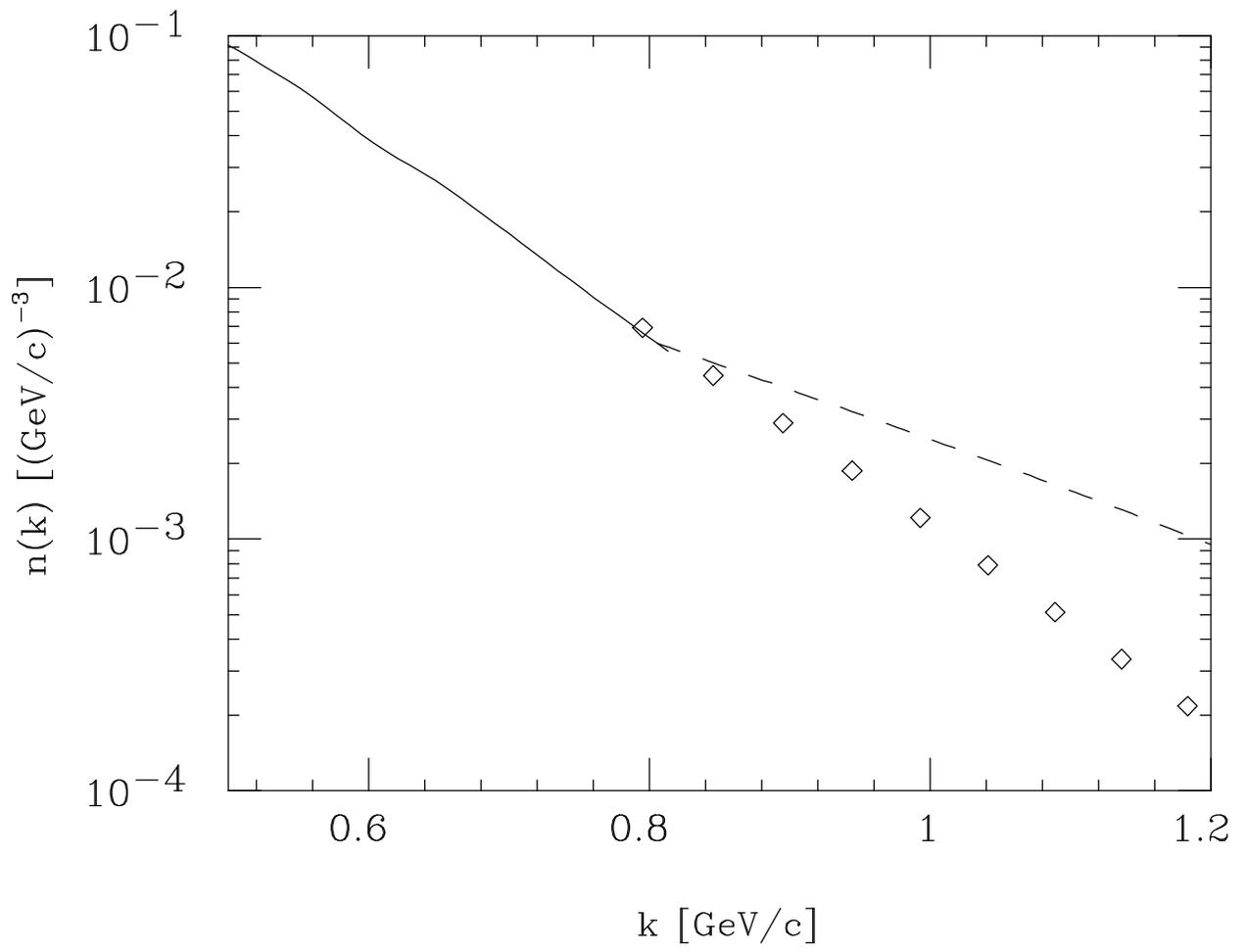
\end{document}